\documentclass[aps,twocolumn,showpacs,superscriptaddress,floatfix]{revtex4}
\usepackage{graphicx}
\usepackage{psfrag}
\begin{document}

\title{The Topological Non-connectivity Threshold \\
in quantum  long-range interacting spin systems}

\author{F.~Borgonovi}
\affiliation{Dipartimento di
Matematica e Fisica, Universit\`a Cattolica, via Musei 41, 25121
Brescia, Italy}
\affiliation{ I.N.F.N., Sezione di Pavia, Italy}
\author{G.~L.~Celardo}
\affiliation{Dipartimento di
Matematica e Fisica, Universit\`a Cattolica, via Musei 41, 25121
Brescia, Italy}
\author{R.~Trasarti-Battistoni}
\affiliation{Dipartimento di
Matematica e Fisica, Universit\`a Cattolica, via Musei 41, 25121
Brescia, Italy}

\begin{abstract}

Quantum characteristics of the Topological 
Non-connectivity Threshold (TNT),
introduced in 
F.Borgonovi, G.L.Celardo, M.Maianti, E.Pedersoli,
J. Stat. Phys.,  116,  516 (2004),
 have been analyzed in the hard quantum regime.
New interesting perspectives in term of the possibility
to study the intriguing quantum-classical transition through 
Macroscopic Quantum Tunneling have been addressed.

\end{abstract}
\date{\today}
\maketitle

\section{Introduction}

The magnetic properties of materials are usually described in the frame
of system models, such as Heisenberg or Ising models where rigorous results,
or suitable mean field approximations are available in the 
thermodynamical limit.
On the other side, modern applications require to deal with  nano-sized
magnetic  materials, whose intrinsic features lead, from one side
to the emergence of quantum phenomena\cite{chud}, and to the other
to the question of applicability of statistical mechanics.
Indeed, few particle systems do not usually fit in the class of systems
where the powerful tools  of statistical mechanics can be applied 
at glance. In particular, an exhaustive theory able to fill the gap
between the description of  $2$ and $10^{23}$  interacting particles
is still missing.
Moreover, also important well-established thermodynamical concepts
as the temperature, become questionable at the nano-scale\cite{tempe}.

In a similar way, long-range interacting systems belong,
since long, in the class where standard statistical mechanics cannot
be applied {\it tout court}. 
Indeed, they  display a number
of bizarre behaviors, to quote but a few, ensemble inequivalence\cite{ruffo},
negative specific heat, temperature jumps and long-time relaxation (quasi-stationary states)\cite{ruf1}.
Therefore, from this  point of view, few-body short-range interacting
systems share some similarities with many-body long-range ones.

Within such a scenario, and thanks to the modern computer capabilities,
it is quite natural take a different point of view,  starting
investigations directly
from the dynamics, classical and quantum as well.
It was in this spirit that, few years ago, a topological non-connection
of the phase space was discovered\cite{jsp} in a class of anisotropic spin systems.
This was initially called, for historical reasons\cite{palmer}, 
breaking of ergodicity, meaning with that a trivial 
consequence, namely 
that the system can not  be ergodic (the phase space is exactly decomposable
in two unconnected parts)\cite{khinchin}, even if we 
prefer here to call it Topological Non-connectivity Threshold (TNT).
This  result,  was found first numerically and later  analytically,
in a class of models, the anisotropic Heisenberg models, where important
and rigorous results have been obtained 
during the last century, in the 
thermodynamical limit only.

Quantum effects in such small  magnetic systems  can not be neglected,
in principle, even if the usual viewpoint \cite{chud}
is to consider magnetic domains as quantum objects with huge spin 
number.
Still,  what we have in mind now and in our future plans,
is to show the relevance of the TNT
with respect to  the complicated
transition between the classical and quantum world.
For instance, it is well known that quantum particles can tunnel across
potential barriers at variance with the classical ones. What is less
obvious is that a macroscopic variable, such as the magnetization,
can do the same. This phenomenon, known as Macroscopic Quantum Tunneling,
well described in \cite{chud} is an important step in the so-called
Leggett program \cite{leggett}
for a better comprehension of
the classical-quantum transition.
Thus, after a brief description 
of the quantum analogue of the classical TNT,
we show its relevance in
single-spin models used in micromagnetism, featuring 
the TNT as a perturbative threshold. 

\section{The Quantum Topological Non-connectivity Threshold}

The  results found in the classical model
\cite{jsp, firenze, brescia}  has been 
considered  in the semiclassical regime in \cite{bcb}.

Here, we consider a system of $N$ particles of spin $l=1$,
described by the following Hamiltonian:
\begin{equation}
\label{eq:quant_ham}
\hat {H} =\frac{\eta}{2} \sum_{i=1}^N \sum_{j\ne i} \hat{S}_i^x 
\hat{S}_j^x 
-\frac{1}{2} \sum_{i=1}^N \sum_{j\ne i} \hat{S}_i^y \hat{S}_j^y,
\end{equation}
where $-1 < \eta \leq 1 $ is the anisotropy constant.
Quantization of the Hamiltonian follows the standard rules.
(Let us remember that, according to the correspondence principle, the
classical limit is recovered as $l \rightarrow \infty$).
As in the classical case we fix the modulus
of the spins to one. This can be achieved with
an appropriate rescaling of the 
Planck constant,  $\hbar \to \hbar /|S_i|=1/\sqrt{l(l+1)}$.
With this choice, in the classical limit, $l \rightarrow \infty$
($\hbar \to 0$),  
the spin modulus remains equal to $1$.
We will also limit our analysis in the subspace of 
all possible completely
symmetric states (bosonic symmetry).

In \cite{bcb}  it was shown that the magnetization along the easy axis,
at variance with the classical case,
can change its sign below the TNT through
Macroscopic Quantum Tunneling.
This leads to the  problem of a significant  definition 
of the quantum TNT.
In the semiclassical limit
(large $l$) a quantum signature of the classical TNT can be found\cite{bcb}
in the spectral properties of
the system leading to a proper  definition of  the 
quantum disconnection threshold, $E_{tnt}^q$,
with the correct classical limit.
Below  $E_{tnt}^q$ the spectrum is characterized by the
presence of quasi degenerate doublets,
whose energy difference, $\delta$,
increases exponentially up to $E_{tnt}^q$, and
saturates  above $E_{tnt}^q$.

\begin{figure}
\includegraphics[scale=0.33]{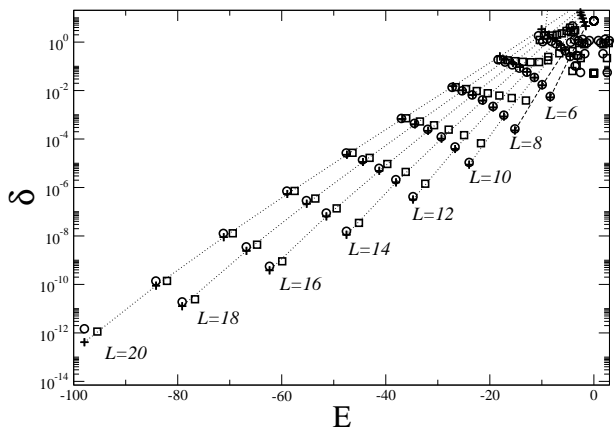}
\caption{
Energy splittings  $\delta$ versus $E$ for
the many--spin case. Eigenvalues and  splittings are compared
for the case $N=20$,  $\eta=1$ and $l=1$.
Circles :   mean field approximation;
squares :  full Hamiltonian;
crosses : perturbative result.
Eigenvalues are arranged in regular block, according
to different  $L$ values as indicated in the figure.
}
\label{deltae}
\end{figure}

On the other side here,   we focus 
on the hard quantum regime ($l=1$).
The energy spectrum, still presents doublets 
and an approximate exponential 
dependence of $\delta$ with the energy.
Nevertheless, it is  evident that,
at variance  with the semiclassical case, they change regularly 
by many order of magnitude in small energy bins, see Fig.~\ref{deltae}.

\begin{figure}
\includegraphics[scale=0.34]{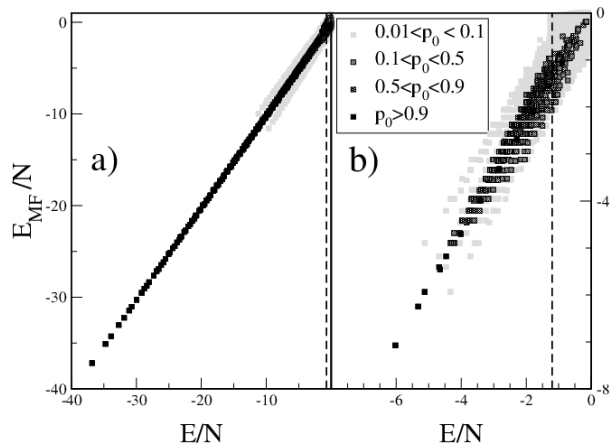}
\caption{
Probability $p_0=|\langle E|E_{MF}\rangle|^2$ 
that an eigenstate $|E\rangle$, 
with specific energy $E/N$, occupies an eigenstate $|E_{MF}\rangle$,
 with specific energy $E_{MF}/N$.
Parameters are  :
a)  $N=100$, $l=1$ ;  b)  $N=6$ $l=4$. 
}
\label{hj}
\end{figure}

In order to understand the origin of this regularities,
it is useful to  rewrite (\ref{eq:quant_ham}) as $\hat{H}=\hat{H}_{MF}+\hat{H}_1$,
where
\begin{eqnarray}
\label{eq:MF}
\hat{H}_{MF} &=& \frac{\eta}{2} \hat{m}_x^2 - \frac{1}{2}  \hat{m}_y^2\\ 
\label{eq:NMF}
\hat{H}_1 &=&  \frac{1}{2}  \sum (\hat{S}_i^y)^2- \frac{\eta}{2}\sum (\hat{S}_i^x)^2. 
\end{eqnarray}

and $\hat{m}_{x,y,z} = \sum_i \hat{S}_i^{x,y,z}$.
While the first term (mean field) 
is integrable in the classical limit,
the  latter   is  responsible for
the non integrability of the system.
Let us also consider the eigenvectors of $\hat{H}_{MF}$, $|E_{MF} \rangle$, and 
expand the eigenvectors of $\hat{H}$, $|E\rangle$ over them. In other words
 we consider the  probability, $p_0 = |\langle E_{MF}|E\rangle|^2$,
 that a given eigenvectors of $\hat{H}$
occupies a given eigenvector of $\hat{H}_{MF}$.
As one can see, see Fig.~\ref{hj}a, 
the eigenvectors of the  full Hamiltonian are almost completely localized on the 
eigenvectors of the mean field Hamiltonian.
over the whole  energy range. 
Actually in the low
energy region the eigenvectors occupies just one eigenstate
of the mean field Hamiltonian with probability greater the 
$0.9$, while all the other states are occupied with 
probability smaller the $0.01$.
The same does not happen in the large $l$ case  (Fig.~\ref{hj}b).
Therefore, the non-integrable part  is negligible
with respect to the mean field (\ref{eq:MF}). 
The question of the quantum
integrability of chaotic Hamiltonians for bosons with $l=1$ has been
recently posed in \cite{benet}. Shortly, quantum integrability should
be induced, for $l=1$, by the strong correlations between Hamiltonian
matrix elements.
From the above analysis it follows that we can use the mean field
Hamiltonian to study the total Hamiltonian in the
hard quantum regime.

Here, we  present the results of a high order perturbative
calculation of the eigenvalues of  $\hat{H}_{MF}$.
Since $[\hat{H}_{MF}, {\hat{m}^2}] =0$, it is sufficient to consider the
eigenvalues of $\hat{m}^2$.
They  are given by the possible values
of the total magnetic moment which can be obtained combining
$N$ particles of spin $l$, and are determined by
the quantum numbers: $Nl, Nl-1,....,0$.
From these values we should exclude those which 
cannot be combined to give completely symmetric states, if
one is interested in the bosonic case (even if 
the present approach is independent
from the statistics). 
We can  consider each subspace with different $m^2$ separately.
In this way the many-spin Hamiltonian $\hat{H}_{MF}$,
 is equivalent to a set of
single spin systems, described by the same Hamiltonian. 
Note that $l$ is the magnitude of the spin of the many-spin problem.
Thus in the following we will first consider single spin models,
and then we will come back to our many-spin problem.

\section{Single Spin Model}
Let us  consider 
a single spin of magnitude 
$L$,  $z-$ component $L_z$, with $|L_z| \le L$.
For simplicity we will rewrite the mean field Hamiltonian as follows:
\begin{equation}
\hat{H}_{MF} = \frac{1}{2}\left( -\hat{m}_z^2+ \eta \hat{m}_x^2 \right)
\label{model1}
\end{equation}
Eq.~(\ref{model1}) can be reduced to (\ref{eq:NMF})
after a rotation of $\pi$ around the $x$ axis which
carries $y$ in $z$ and $z$ in $-y$, and does not  
affect the physics of the problem.

Single-spin-model have an interest in themselves,
besides the fact that their analysis will allow us to
compute the energy levels of our many-spin mean field Hamiltonian.
In recent years growing interest arose in micromagnetic particles
\cite{chud,takagi}, such as ferromagnetic 
domains  and magnetic  macro--molecules such as $Mn_{12}$ and $Fe_8$
\cite{molecule}.  
The research interest in these systems is mainly due  
to the possibility to reveal quantum effects
in the macroscopic domain,
such as the Macroscopic Quantum Tunneling (MQT) of the
magnetic moment, and the 
even more interesting phenomenon of Macroscopic Quantum Coherence (MQC).
While in the former case (MQT) the total magnetization of a microscopic
particle flip even if classically this would be forbidden
by the presence of an effective energy barrier, in the latter case (MQC)
the magnetization oscillates between opposite magnetization states
in a coherent way. This phenomenon, if revealed, would 
unambiguously indicate the presence of  Quantum Interference of
Macroscopic Distinct States \cite{leggett}.
At sufficiently low energy these systems
can be modeled by phenomenological single--spin Hamiltonians, 
where the single spin describes the
total magnetic moment of the system.
Splittings of the eigenvalues 
of the single spin Hamiltonians are simply related 
to the frequencies of MQC (or the tunneling rates of MQT) \cite{chud}. 
For this reason much effort has been devoted in these years to 
compute such splittings. 
Usually,  semiclassical methods are employed, such as WKB
and imaginary time path integrals to quote but a few \cite{svizzeri,chud2}.
Also perturbation theory can be successfully
applied in this kind of problem taking into account high order terms
\cite{garanin}.
Indeed it is possible to compute
explicitly 
the first non--zero perturbative contribution 
to the splittings even if 
this is an high--order contribution.

In Appendix \ref{pert} we show the basic ideas of the 
high  perturbative order approach, and we derive an analytical
expression for the eigenvalues and the corresponding splittings
(in \cite{garanin} no explicit derivation
was given).

\begin{figure}
\includegraphics[scale=0.34]{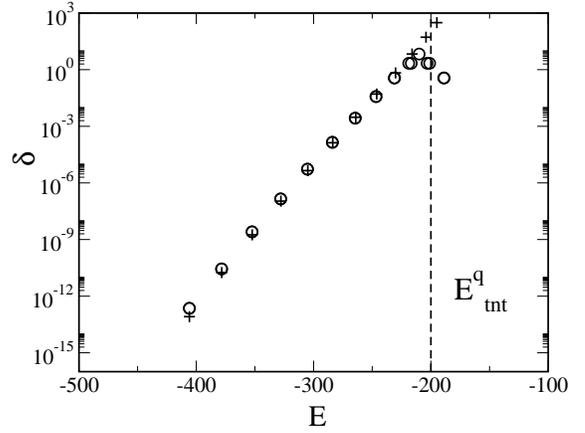}
\caption{$\delta$ shifts versus the energy $E$.
Eigenvalues and  splittings  obtained
numerically (open circles) 
and via perturbation theory  (crosses)  for a single spin $L=20$.
Also shown as vertical dashed line 
 the rough estimates for the validity of
the our approach.
Here is  $\hbar=1$, $\eta=-0.5$.}
\label{etashift}
\end{figure}

Numerical eigenvalues and their splittings
for  Hamiltonian (\ref{model1}) and  a single 
spin of magnitude $L$, have been compared with  our results in
Fig.~\ref{etashift},
where  we show the splittings $\delta$ as a
function of the energy $E$.
In the same figure
we can see that while the high perturbative order 
approach gives a very good estimate for $E < E^q_{tnt}$
it fails completely above the threshold. 
An upper bound to the energy at which 
our approach fails can be given evaluating the
quantum correction  to $E_{tnt}$ for the   
single spin Hamiltonian, see \cite{bcb} (indicated  in Fig.~\ref{etashift} 
as a vertical dashed line).

\section{Many spin Hamiltonian}

We  now compare our analytical results  
with the eigenvalues of the mean field Hamiltonian $H_{MF}$
considered as a many spin Hamiltonian and
with the eigenvalues of the complete Hamiltonian
in the hard quantum regime (\ref{eq:quant_ham}).
We can achieve this by considering the 
possible eigenvalues of $\hat{m}^2$, which are 
obtained when an ensemble of $N$ particles of 
spin $l$ is considered.
Note that from the set of possible eigenvalues of $\hat{m}^2$
we have to exclude those
that are not compatible
with the symmetrization postulate (for the bosonic case).
For each possible value of $\hat{m}^2$ we apply our 
perturbative approach to the correspondent single spin
problem. Then, putting all  together, we obtain 
the  results for the many--spin Hamiltonian.

In Fig.~\ref{deltae} we plot the splittings versus the 
specific energy for the mean field Hamiltonian (\ref{eq:MF}) (circles),
for the full Hamiltonian (\ref{eq:quant_ham}) (squares),
and the perturbative results (crosses). 
As one can see we can give an accurate  good approximation in the low energy region
of the spectrum.
Deviations obviously appear when the perturbative approach
is compared with the splittings of the full Hamiltonian,
even if,  in this case, the perturbative approach still give a 
correct order of magnitude 
estimate of  eigenvalues and splitting.
 From Fig.~\ref{deltae} we can also see how the regular 
features of $\delta$  in the hard quantum regime 
are related to the quantum numbers of the total angular momentum.

One could ask if the same perturbative approach  works
in the semiclassical regime for the total Hamiltonian (\ref{eq:MF}).
Perturbation theory cannot work in the 
whole  energy region, even if we may expect to give an approximate
description  for low--energy eigenvalues. For instance, applying perturbation theory,
e.g. Eq.~(\ref{delta}),
for  the energy separation between the ground state and the first 
excited state, we get:
$ \delta_{GS} = 0.25  \hbar^2/ 6^{-Nl}$.
This expression works 
works fairly well\cite{celardo}, even  for large $l$ and it 
reproduces the main features of 
the dependence of the ground state splitting, namely 
the exponential dependence on 
$l$, and  $N$ as well.

We have shown that, in the hard quantum regime, it is possible
   to compute perturbatively the splittings of the doublets
  characterizing the spectrum below the quantum TNT.
   This is due to the nearness  of the full Hamiltonian with
   the mean field Hamiltonian.
   The quantum TNT can be also considered a perturbative threshold, since it
   gives an estimate of the energy at which the perturbative
   approach fails.
   Finally, we  point out that this threshold indicates
an energy range
which is not negligible with respect to the total energy range for long range 
interacting systems
in which Macroscopic Quantum Phenomena can be studied.

\section{Appendix}
\label{pert} 

In this section we  present the results of a high  perturbative order
calculation of the eigenvalues of the single spin Hamiltonian
(\ref{model1}).
We will show that in order to split the double degenerate
levels of the $-\hat{m}_z^2$ term of (\ref{model1})  
characterized
by the quantum number $l_0=|m_z|$, the first non--zero contribution
is at the $l_0$-th perturbative order.
This also give a qualitative explanation of the well known exponential
dependence of the splitting magnitudes on the energy.

Let us consider a single spin of magnitude $L$,  $z$-component 
$L_z$ with $|L_z| \le L$ and $|L,L_z\rangle$ as basis states.
Hamiltonian (\ref{model1})
can be written as $\hat{H}=\hat{H}_0+\hat{V}$, where~:
\begin{eqnarray}
\hat{H}_0&=& -\hat{m}_z^2+ \frac{1}{4} \eta (\hat{m}^+\hat{m}^-+\hat{m}^-\hat{m}^+) \nonumber \\
\hat{V}&=&\frac{1}{4} \eta (\hat{m}^+\hat{m}^++\hat{m}^-\hat{m}^-)
\label{eq:h0v}
\end{eqnarray}
and $\hat{m}^{\pm}=\hat{m}_x\pm i\hat{m}_y$.
Since $\hat{H}_0$ is diagonal in the basis $|L,L_z\rangle$,
the unperturbed energy $E_0=\langle L,l_0 |\hat{H}| L,l_0\rangle$ 
are given by:
\begin{equation}
E_0(L,l_0)= - \hbar^2 l_0^2 + \frac{1}{2} \hbar^2 \eta \left[ L(L+1)-l_0^2)
\right]
\label{E0}
\end{equation}
Each unperturbed energy  level
turns out to be  doubly degenerate, with eigensubspaces spanned by 
$|L, \pm l_0 \rangle$.
The first non-zero contribution
to the splitting of 
a degenerate  pair  occurs  at the $l_0$-th order of perturbation theory.
In order to show that 
let us define the $n-$th order perturbation operator \cite{sakurai}:
\begin{equation}
\hat{\cal P}^{(n)}= \hat{V} \left(\frac{\hat{\phi}}{E_0-H_0} \hat{V}\right)^{n-1},
\label{eq:P}
\end{equation}
 where 
$\hat{\phi} =1-\sum_{{E'}_0 \ne E_0} |{E'}_0 \rangle \langle {E'}_0|$ 
is the projector out of the considered degenerate subspace.
The right linear combination of the unperturbed
basis vectors $|L, \pm l_0\rangle$ (to which eigenstates of 
$\hat{H}$ tend when  $\hat{V}$ is negligible) can be found 
by diagonalizing  the following matrix:
\begin{equation}
\pmatrix{{\cal P}_{++}^{(n)} & {\cal P}_{+-}^{(n)} \cr 
{\cal P}_{-+}^{(n)} & {\cal P}_{--}^{(n)}}
\label{eq:matrix2}
\end{equation}
where ${\cal P}_{s s'}^{(n)}=
\langle L,sl_0|\hat{{\cal P}}^{(n)}| L,s'l_0 \rangle$,
$s,s'=\pm1$
and $n$ is the minimum  order 
giving rise  to  two different
eigenvalues of the matrix (\ref{eq:matrix2}).
A $\pi-$rotation around the $x-$axis
leaves $\hat{V}$ unchanged since 
$|L,l_0 \rangle \to |L,-l_0 \rangle$,
and $\hat{m}^\pm \rightarrow \hat{m}^\mp$. Then
$ {\cal P}^{(n)}_{++}= {\cal P}^{(n)}_{--}$, 
$ {\cal P}^{(n)}_{+-}= {\cal P}^{(n)}_{-+}$
and the right combination of 
unperturbed basis vectors is:
\begin{equation}
\label{eq:Del3}
|\pm m_0 \rangle =  ( |L, l_0 \rangle \pm  |L, -l_0 \rangle)/\sqrt{2}.
\end{equation}
Eigenvalues undergo a shift  given by :
$\pm {\cal P}_{+-}^{(n)}$.
The generic $n$-th order energy shift, $\Delta^{(n)}$ induced by the 
perturbation is given by:
$\Delta^{(n)}= \langle \pm m_0 | \hat{\cal P}^{(n)} | \pm m_0 \rangle$.
While the degeneracy can be removed only by non-zero off-diagonal
elements, an overall energy shift $D$ can be induced by 
non-zero diagonal elements too.

In order to compute $D$ and $\delta$ the action of $\hat{\cal P}^{(n)}$
on the basis states $|L,\pm l_0\rangle$ should be evaluated.
If $n=1$ then $\hat{\cal P}^{(1)}=\hat{V}$. In this case the diagonal elements of the 
matrix (\ref{eq:matrix2}) are zero since
$\hat{V}$ can only change $l_0\to l_0 \pm 2$.
Off--diagonal elements 
$\langle L,-l_0|\hat{V}|L,+l_0\rangle $
are different from zero  only when 
$\hat{V}$ brings $|L, l_0\rangle$ into $|L,- l_0\rangle$. 
This can happen only if  $l_0=1$.

If $n=2$ then 
$\hat{\cal P}^{(2)}=\hat{V} (\hat{\phi}/(E_0-H_0) ) \hat{V} $.
Since  
\begin{equation}
\frac{\hat{\phi}}{(E_0-H_0)} |L,l \rangle = 
\frac{1-\delta_{l,l_0}}
{ E_0(L,l_0)-E_0(L,l) } |L,l \rangle ,
\end{equation}
in order to understand the action of $\hat{\cal P}^{(2)}$ we have to apply 
$\hat{V}$ twice.
Let's consider the diagonal elements:
Can we take $|L,l_0\rangle$ in itself $|L,l_0\rangle$, using $\hat{V}$ twice?
Yes: 
$\hat{V}\hat{V} |L,l_0\rangle \rightarrow \hat{V}(|L,l_0-2\rangle + |L,l_0+2\rangle) 
\rightarrow |L,l_0\rangle  +|L,l_0-4\rangle  +|L,l_0+4\rangle + |L,l_0\rangle$,
where the coefficients in front of the states have been omitted.
Bracketing the final states thus obtained with $|L, l_0\rangle$, 
only the first and the last remain.
Then there are two ``ways'' in which the
operator $\hat{\cal P}^{(2)}$ can take $|L,l_0\rangle$ in itself:
if $l_0 > 1$ by the following chain rule :
$|L,l_0\rangle \rightarrow |L,l_0-2\rangle \rightarrow |L,l_0\rangle $, 
while  if $l_0<L-1$ by  
$|L,l_0\rangle \rightarrow |L,l_0+2\rangle \rightarrow |L,l_0\rangle $.

It is now easy to compute the first non zero contributions
to the overall shift. From (\ref{eq:Del3}) we have:
\begin{equation}
\label{eq:D}
D= \langle L, \pm l_0|\hat{V} \left(\frac{\hat{\phi}}{E_0-H_0} \hat{V}\right)|L, \pm l_0 \rangle,
\end{equation}
The only contributions $D_{\pm}$ different from zero, coming from the
two ways described above, are:

\begin{equation}
\label{DD}
D_{\pm}=
\left(\frac{\displaystyle \eta \hbar}{\displaystyle 4}\right)^2 
\frac{\displaystyle f^{\pm}(L,l_0) f^{\pm}(L,l_0-1)}{\displaystyle \Delta E_0^{\pm}},
\end{equation}
respectively for $l_0 >1$ ($D_-$) and $l_0 < L+1$ ($D_+$). Here, we defined 
$f^{\pm} (L,l_0) = L(L+1) -l_0(l_0\pm 1)$
and $\Delta E_0^\pm = E_0(L,l_0)-E_0(L,l_0 \pm 2)$.
Thus, $D=D_-+D_+$ is the first non--zero overall energy shift.

Let us now  consider  the off-diagonal matrix elements.
It is possible to go from $|L,l_0\rangle$ to $|L,-l_0\rangle$ using $\hat{V}$ twice
only when  $l_0=2$.
In this case there is one only way:
$|L,l_0\rangle \rightarrow |L,l_0-2\rangle \rightarrow 
|L,l_0-4 \rangle =
|L,-l_0\rangle $, the last being true only for $l_0=2$.
It is then clear 
why the first non-zero  operator which splits
the doublet characterized by  $L,l_0$ is the $l_0$-th order.
 From (\ref{eq:Del3}) we have:

\begin{equation}
\delta=2 \langle L,l_0|\hat{V} \left(\frac{\hat{\phi}}{E_0-H_0} \hat{V}\right)^{l_0-1}
|L,-l_0 \rangle
\label{d1}
\end{equation}

 From Eq.~(\ref{d1}) there is only one way to connect 
$|L,l_0 \rangle$ with  $|L,-l_0 \rangle$, namely 
$|L,l_0 \rangle \rightarrow |L,l_0-2 \rangle ......\rightarrow |L,-l_0+2 \rangle\rightarrow |L,-l_0 \rangle $. 
After some algebra, one has: 

\begin{equation}
\delta= \hbar^2 \left(\frac{\eta}{4}\right)^{l_0} (-1)^{l_0-1} 
\frac{\prod_{j=-l_0}^{l_0-1} \sqrt{L(L+1)-j(j+1)} }{\prod_{p=1}^{l_0-1} (4+2 \eta)p(l_0-p) }
\label{delta}
\end{equation}

To summarize, for any  given degenerate doublet 
we can calculate the overall shift $D$, 
Eqs.(\ref{DD}) 
and the splitting $\delta$, Eq.(\ref{delta}).

\end{document}